\documentclass[9pt,conference]{IEEEtran}
\IEEEoverridecommandlockouts

\usepackage{cite}
\usepackage{amsmath,amssymb,amsfonts}
\usepackage{algorithmic}
\usepackage{graphicx} 
\graphicspath{{images/}} 

\usepackage[hyperfootnotes=false]{hyperref}
\usepackage{amsmath}
\usepackage{tikz}
\usepackage{pgfplots}
\usepackage{textcomp}
\usepackage{xcolor}
\pgfplotsset{compat=1.18}

\newcommand\blfootnote[1]{%
  \begingroup
  \renewcommand\thefootnote{}\footnote{#1}%
  \addtocounter{footnote}{-1}%
  \endgroup
}

\def\BibTeX{{\rm B\kern-.05em{\sc i\kern-.025em b}\kern-.08em
    T\kern-.1667em\lower.7ex\hbox{E}\kern-.125emX}}
\begin{document}

    \title{Ulcerative Colitis Mayo Endoscopic Scoring Classification with Active Learning and Generative Data Augmentation
}

\author{\IEEEauthorblockN{Ümit Mert Çağlar*}
\author{
    \IEEEauthorblockN{
        Ümit Mert Çağlar\IEEEauthorrefmark{1}\IEEEauthorrefmark{2}, Alperen İnci\IEEEauthorrefmark{1}\IEEEauthorrefmark{2}, Oğuz Hanoğlu\IEEEauthorrefmark{2}, Görkem Polat\IEEEauthorrefmark{2}, Alptekin Temizel\IEEEauthorrefmark{2}\IEEEauthorrefmark{3}
    }
        \IEEEauthorblockN{
        \{mert.caglar,
alperen.inci,
oguz.hanoglu,
gorkem.polat,
atemizel\}@metu.edu.tr
    }
    \IEEEauthorblockA{\IEEEauthorrefmark{2} \textit{Graduate School of Informatics} \textit{Middle East Technical University}}
    \IEEEauthorblockA{\IEEEauthorrefmark{3} Neuroscience and Neurotechnology Center of Excellence (NOROM) \\
Ankara, Turkey}
}

\IEEEauthorblockA{\textit{Graduate School of Informatics} \\
\textit{Middle East Technical University}\\
Ankara, Turkey \\
mert.caglar@metu.edu.tr}
\and
\IEEEauthorblockN{Alperen İnci*}
\IEEEauthorblockA{\textit{Graduate School of Informatics} \\
\textit{Middle East Technical University}\\
Ankara, Turkey \\
alperen.inci@metu.edu.tr}
\and
\IEEEauthorblockN{Oğuz Hanoğlu}
\IEEEauthorblockA{\textit{Graduate School of Informatics} \\
\textit{Middle East Technical University}\\
Ankara, Turkey \\
oguz.hanoglu@metu.edu.tr}
\and
\IEEEauthorblockN{Görkem Polat}
\IEEEauthorblockA{\textit{Graduate School of Informatics} \\
\textit{Middle East Technical University}\\
Ankara, Turkey \\
gorkem.polat@metu.edu.tr}
\and
\IEEEauthorblockN{Alptekin Temizel}
\IEEEauthorblockA{\textit{Graduate School of Informatics} \\
\textit{Middle East Technical University}\\}
\IEEEauthorblockA{\textit{Neuroscience and Neurotechnology Center of Excellence (NOROM)} \\
Ankara, Turkey \\
atemizel@metu.edu.tr}
}
\IEEEoverridecommandlockouts
\IEEEpubid{\makebox[\columnwidth]{979-8-3503-3748-8/23/\$31.00~\copyright2023 IEEE \hfill} \hspace{\columnsep}\makebox[\columnwidth]{ }}
\maketitle
\IEEEpubidadjcol
\blfootnote{{\IEEEauthorrefmark{1}Equal contribution}}
\begin{abstract}
 Endoscopic imaging is commonly used to diagnose Ulcerative Colitis (UC) and classify its severity. It has been shown that deep learning based methods are effective in automated analysis of these images and can potentially be used to aid medical doctors. Unleashing the full potential of these methods depends on the availability of large amount of labeled images; however, obtaining and labeling these images are quite challenging. In this paper, we propose a active learning based generative augmentation method. The method involves generating a large number of synthetic samples by training using a small dataset consisting of real endoscopic images. The resulting data pool is narrowed down by using active learning methods to select the most informative samples, which are then used to train a classifier. We demonstrate the effectiveness of our method through experiments on a publicly available endoscopic image dataset. The results show that using synthesized samples in conjunction with active learning leads to improved classification performance compared to using only the original labeled examples and the baseline classification performance of 68.1\% increases to 74.5\% in terms of Quadratic Weighted Kappa (QWK) Score. Another observation is that, attaining equivalent performance using only real data necessitated three times higher number of images.
 
\end{abstract}


\section{Introduction}
Endoscopic images play a crucial role in the diagnosis and treatment of various gastrointestinal diseases. However, acquisition of endoscopic images is a costly process and necessitates careful consideration of data and patient privacy-related issues; furthermore, the process of labeling is time-consuming and costly, as it requires the expertise of trained medical professionals. A reliable labeling process involves blind labeling by multiple medical professionals and resolving labeling inconsistencies to obtain accurate annotations \cite{limucpaper2022}. Deep learning based classification methods typically require a large number of labeled examples to achieve good performance and, in practice, it may not be feasible to manually label a sufficient number of endoscopic images, particularly for rare or subtle abnormalities. Our empirical findings, yielding a QWK score of 74.5\%, demonstrate comparable performance to the classification proficiency exhibited by human experts \cite{qwk_expert_comparison}.

Data augmentation enhances the performance of data-driven models by intentionally altering input data during the training process. In addition, synthetically generated images can also be used to increase the number of labeled images in the dataset, while these images might not always contribute to the model performance if they are very similar to the existing ones or if they are out-of-distribution. 

Active learning is a machine learning approach which aims to select the most informative examples to label and subsequently use in training of a classifier. By actively selecting the most informative examples, it is possible to achieve good performance with a smaller number of labeled examples. In this paper, we propose a method for synthesizing additional data and further selection of these using active learning for endoscopic image classification. 

\section{Related Work}
Ulcerative Colitis (UC) is a chronic inflammatory bowel disease that affects the colon and rectum. In this section, first, related works on the automatic classification of UC are provided. When there is a limited number of labeled images, deep generative models can be used to improve the model performance by generating and adding synthetic examples. So, next, such models that could be used to synthesize data are discussed. Then, active learning methods, which could assist in determining how to pick the synthetic samples that will be most helpful throughout the training are summarized.

\subsection{Ulcerative Colitis Classification }
\begin{figure}[ht]
    \centering
    \begin{tabular}{c c c c}
        \textbf{{MES 0}} & \textbf{{MES 1}} & \textbf{{MES 2}} & \textbf{{MES 3}}\\
        healthy & mild & moderate & severe\\
        \includegraphics[width=0.10\textwidth]{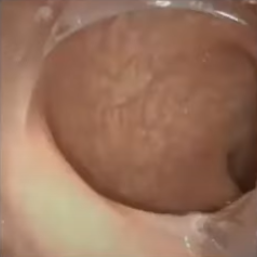} & 
        \includegraphics[width=0.10\textwidth]{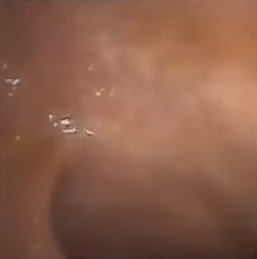} &
        \includegraphics[width=0.10\textwidth]{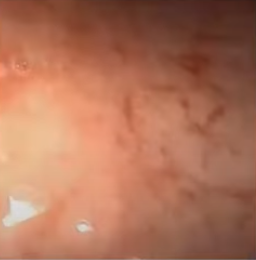} & 
        \includegraphics[width=0.10\textwidth]{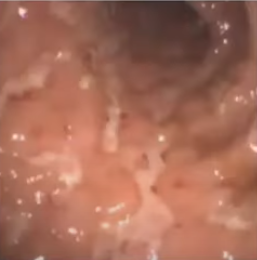}
        
    \end{tabular}
    \caption{Real endoscopy images corresponding to different MES levels.}
    \label{fig:mayo_images}
\end{figure}

Accurate classification of UC severity is important for proper diagnosis and treatment, as the disease can result in serious complications if left untreated. Traditionally, UC classification has been performed by trained medical professionals based on endoscopic images and other clinical data. However, manual classification can be subjective and time-consuming and may not be feasible for large datasets. Mayo Endoscopic  Scoring (MES) system involves categorizing endoscopic images to four distinct severity levels of the disease. The scoring is based on several factors, including mucosal bleeding and ulceration. Sample images corresponding to different MES levels are shown in Fig. \ref{fig:mayo_images}.

Deep learning models have the potential to significantly improve the efficiency and accuracy of UC classification, and can be used to support medical professionals in their decision-making process ~\cite{AIinIBDreview2021}. These models are trained to learn the underlying patterns in endoscopic images and other clinical data, and can classify ulcerative colitis with high accuracy. 

\subsection{Generative Adversarial Networks }
Generative Adversarial Networks (GAN) allow synthesizing data that is similar to an original dataset~\cite{Goodfellow2014} distribution. StyleGAN2, introduced in 2020, was shown to be able to generate high-resolution images~\cite{Karras2020a}. StyleGAN2-ADA is an extension over the StyleGAN2 approach and introduces adaptive discriminator augmentation, which applies various augmentations to the images, where the discriminator component of the network observes and adaptively changes the frequency and amplitude of the augmentation~\cite{Karras2020a}. It is shown that adaptive discriminator augmentation allows training GAN models with limited data.

\subsection{Active Learning }
Active learning is a machine learning technique that involves actively selecting the most informative examples to label, which are then used to train a classifier. In this paper, active learning methods are adapted to work in a setting where they are used to select informative samples from a synthetically generated data pool for endoscopic image classification.

Main active learning approaches are diversity-based methods, entropy-based methods, and margin-based methods. 
	
\paragraph {Diversity-based methods} 
\textit{Diversity}-based methods seek to identify a subset which is representative of the entire dataset. This subset can be used to train a classifier, while minimizing the number of labeled examples required. The \textit{Coreset} technique \cite{coreset2018} looks for a diverse set of points with the maximum distance from others to represent the whole dataset. The method determines the pairwise distances between the labeled set and the unlabeled set, in our case the pool of synthetic data.

\paragraph{Entropy-based methods} 
Entropy-based methods involve selecting examples for labeling based on the uncertainty of the current classifier's prediction probabilities. Specifically, these methods choose examples that are least certain according to the classifier's current prediction probabilities. The intuition behind this approach is that by selecting examples that the classifier is currently uncertain about, we can improve its generalization performance by providing it with additional information about the underlying data distribution. Entropy-based methods are particularly useful when the cost of labeling is relatively high, as they allow the classifier to focus on the most informative samples first. One disadvantage of entropy-based methods is that they may not always select the most ``difficult" examples, as they are based on uncertainty rather than difficulty. However, they have been shown to be effective in a variety of settings and are a popular choice for active learning.

To identify the least certain examples, entropy-based methods calculate the entropy of the classifier's prediction probabilities for each example in the dataset. The example with the highest entropy is selected for labeling, as it represents the most uncertainty in the classifier's predictions. The process is then repeated until a sufficient number of samples have been labeled.

\begin{equation}
    \label{eq:entropy}
    \phi_{E N T}(x)=-\sum_{y} P_{\theta}(y \mid x) \log P_{\theta}(y \mid x)
\end{equation}
		
\paragraph {Margin-based methods} 
Margin-based methods involve selecting examples that are the most ``difficult" to classify according to the classifier's current decision boundary. These examples are typically located near the boundary between different classes and are, therefore, the most informative for improving the classifier's performance. Margin-based methods aim to maximize the margin between the decision boundary and the closest examples from each class, as this results in a classifier that is more robust and less prone to overfitting. 

\section{Proposed Method}
\subsection{Dataset}
We used the LIMUC dataset~\cite{limucpaper2022} which consists of a total of 19,537 raw endoscopic images from 1,043 colonoscopy procedures of 572 UC patients. These images were analyzed by experienced gastroenterologists, resulting in a final dataset of 11,276 images. It is worth noting that the dataset is imbalanced, with the distribution of Mayo Endoscopic Scoring (MES) levels being 54.14\% for the MES-0, 27.07\% for the MES-1, 11.12\% for the MES-2, and 7.67\% for the MES-3.

To ensure the accuracy and validity of our research, we removed 992 endoscopic images that contained instruments within the frame. The presence of instruments could be detrimental for the deep learning model. Additionally, instruments could introduce challenges and problems when training GAN models. Our goal was to obtain synthetic endoscopic images that closely resemble different levels of UC.

Furthermore, we created subsets of the original dataset comprising smaller and balanced datasets. These subsets contained 50 images per class and were incrementally increased up to 500 images per class. We used these subsets to evaluate the baseline performance and compare it with the performance of our proposed method.

\subsection{System Architecture}

\begin{figure}[ht]
	\centering
	\includegraphics[width=0.45\textwidth]{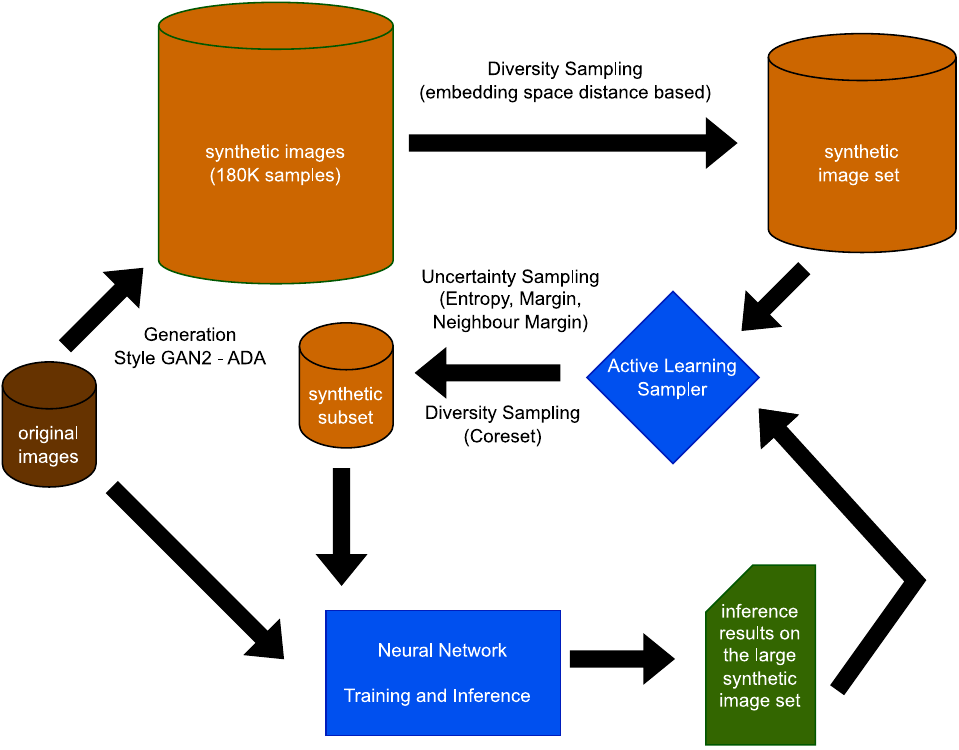}	
	\caption{The System Architecture - Our proposed work involves synthesizing a very large synthetic dataset from the original endoscopy images and using active learning methods to select the most informative samples.}
	\label{fig:system}
\end{figure}

The overall system architecture is shown in Fig. \ref{fig:system}. To generate synthetic colonoscopy images, we used  StyleGAN2-ADA model, which is a state-of-the-art deep generative model for synthesizing high-resolution images. We trained the model on a small dataset of real colonoscopy images and used it to synthesize a large pool of synthetic colonoscopy images. We used PyTorch implementation of \textit{StyleGAN2-ADA}\cite{Karras2020} with a resolution of $256\times256$ since it is highly efficient in limited data use cases. Due to the limitations on class-conditional GAN training, we have not employed transfer learning to these models and trained them from scratch. For class-specific models, we have applied transfer learning and used FFHQ-256\cite{ffhq_pretrained} as the base and continued training GAN models using transfer learning. Our experiments from previous works for class-conditional and class-specific settings prove that class-specific GANs are better in our limited data use case. Class-conditional GANs improved MES classification performance by 4.16\% in best case, while class-specific GANs improved it by 7.05\% in best case. Due to more promising results, we have employed class-specific GANs with transfer learning in this work.

The truncation parameter plays a crucial role in  the diversity of synthetic samples. A value of 0 results in identical generated outputs, while 1 mirrors the diversity seen in the training set during GAN training. When the truncation parameter ranges between 0 and 1, generated samples adhere to the distribution of the training dataset. Values exceeding 1 amplify diversity but can also yield unrealistic outputs. Despite variations in image properties resulting from different truncation values, no optimal choice exists for maximizing classification scores in endoscopic image classification.

In this work, we used the truncation values of 0.5, 1.2 and 2.0. This selection is due to the inherent bias associated with strictly in-distribution samples, introduced by a value of 0.5 as a safety measure. Moreover, a truncation value of 1.2 closely resembles samples with a truncation value of 1.0. The most diverse samples are generated with a truncation value of 2.0. Our approach combines these three values of 0.5, 1.2, and 2.0 to create a large synthetic dataset containing both diverse samples and those resembling the original dataset. 

The utilization of active learning (coreset-based, entropy-based or margin-based methods) methods enables us to eliminate excess images from this extensive dataset. The latest model trained on both original and synthetic images from the last iteration calculates features and active learning scores for the synthetic set. Then, using these scores, the active learning sampler selects the synthetic images to include in the next iteration.

Synthetic samples tend to be unrealistic and can even become outliers, particularly when a small amount of data is used in GAN training. Therefore we propose a method called \textbf{Neighbour Margin}. Neighbour Margin excludes samples for which the calculated distance does not align with the two nearest neighboring classes. This approach takes into account the ordinal nature of the classes (Eq. \ref{eq:neighbour_margin}).

\begin{equation}
\label{eq:neighbour_margin}
    \phi_{NM}(x) = 
    \begin{cases}
        P_{\theta}\left(y_{1}^{*} \mid x\right)-P_{\theta}\left(y_{2}^{*} \mid x\right),& \text{if } |{y_{1}^{*}} - {y_{2}^{*}}| == 1 \\
        \infty,              & \text{otherwise}
    \end{cases}
\end{equation}

$P_{\theta}\left(y_{1}^{*} \mid x\right)$ represents the probability of the model ${\theta}$ on the given sample $x$ for the class $y_1^*$. $y_1^*$ and $y_2^*$ represent the classes with the best and second best probabilities. 
$\phi_{NM}$ represents the neighbour margin based uncertainty score of the sample $x$.
${|{y_{1}^{*}} - {y_{2}^{*}}|}$ is a term that penalizes the samples with class distance larger than 1.

The selected synthetic images were combined with the original labeled examples to form the training set for our classifier. We used ResNet18 \cite{He2016}, a convolutional neural network (CNN), as the classifier, and trained it using the training set.

\subsection{Baseline Image Classification}

ResNet18 deep learning models were trained to classify images according to the MES. The image classification component of our work was optimized by following the strategy described in \cite{polat2022class}. The state-of-the-art training strategies for image classification were used as the baseline to evaluate the performance of our model.

To further evaluate our model's performance, we employed subsets of the original dataset with varying numbers of images per class (50, 100, ..., 450, 500) for four different levels of MES. These subsets were used to assess the saturation level of the image classification network and generate baselines using smaller hypothetical and balanced datasets.

After obtaining the baseline performance using real images, we explored additional methods utilizing synthetic images generated through GAN models trained solely on real images.

\subsection{Generative Data Augmentation}
\paragraph {GAN Training} 

The GAN training process can be monitored by tracking performance metrics such as the Frechet Inception Distance (FID)\cite{FID}. FID is a deep learning-based distance metric that measures the similarity between synthetic images and the original dataset in latent space by comparing hidden layer activations. A lower FID score indicates that the synthetic images are more similar to the originals, which is desired. However, it is important to note that for generative data augmentation, a lower FID score may not directly lead to improved classification performance, as the focus is on generating diverse and realistic images rather than optimizing for classification tasks. 
    
\paragraph {Image Synthesis Tuning}

One of the critical concerns regarding image synthesis with Style-GAN2-ADA is the truncation parameter. Truncation determines the characteristics of the synthetic samples. Values lower than 1 indicate that the generated samples come from the original data distribution; when it is 0, all generated synthetic images will be identical, and when it is 1, the dispersion of the generated data will match that of the training set, values larger than 1 increase the diversity. However, increasing the diversity may result in outliers and unrealistic samples. As there is no single best option for the truncation value, we employed different truncation values of 0.5, 1.2, and 2.0, ensuring balance and variety. By utilizing only 200 real images, we created a diverse synthetic data set comprising a total of 180,000 images.

\paragraph {GAN Synthesized Image Evaluation} 

 GAN models often generate synthetic images that closely resemble the original data. After using the Fiftyone tool \cite{moore2020fiftyone} to visualize the generated images, we noticed that this effect becomes even more pronounced in our work due to the use of a very limited dataset consisting of only 50 images.
 
 To address this similarity issue, we employed diversity-based methods, which involve selecting a representative subset of examples from the dataset. These methods aim to identify the most informative examples for training a classifier. We applied the coreset technique proposed by \cite{coreset2018}, which aims to find a diverse set of points that are maximally distant from one another to represent the entire dataset. In our case, this involved calculating pairwise distances among our synthetic pool. Using the Euclidean distance metric on the image embeddings, we selected the most diverse samples. Consequently, out of our extensive pool of 180,000 synthetic images, we obtained a subset of 1,698 images that exhibited the highest level of diversity.
\subsection{Active Learning}
Active learning methods were employed to expedite the extensive parameter search process and achieve a balance between optimum model performance and time allocation. The proposed technique, neighbour margin, was utilized to select synthetic samples and added to the training set until the model convergence. Once the baseline was obtained, an equal number of samples were incrementally added to four classes. This approach was applied to a subset of the original dataset, containing 50 images per class, with the aim of demonstrating that labeling effort can be reduced by generative data augmentation with active learning.

\section{Experiments}
\subsection{Image Classification}

In this work, we utilized ResNet18 \cite{He2016} as it is a faster and lighter model, making it more suitable for smaller datasets. Our experiments were based on a single network size, and we conducted saturation experiments to determine the point at which increasing the dataset size no longer improved performance. We developed our baseline image classification framework based on the works in \cite{polat2022class}. We have used PyTorch framework and Python to implement experiments in this work. We have followed the official StyleGAN2-ADA implementation \cite{pytorch-ada-github} and monitored our work with the experiment tracking tool Weights and Biases (WANDB) \cite{wandb}.

\subsection{Synthetic Image Generation}

First, we experimented with two distinct approaches: using seperate class-specific GANs for different classses and using a conditional GAN which uses a single model for all four classes with class-based influence on outputs. However, implementation limitations in StyleGAN2-ADA do not allow transfer learning for this approach and we trained separate models for each class, totaling four models, and utilized transfer learning from the FFHQ-256 dataset for these class-specific GANs.

\subsection{Performance Evaluation}

We have used accuracy, precision, recall, F1 score and Quadratic Weighted Kappa (QWK) score for MES classification, as well as FID score for GAN model training. FID score is measured by a neural network by comparing the normal distribution of original and synthetic data through Wasserstein Distance\cite{WassersteinDistance}.

We have conducted a series of experiments to compare the performance of a classifier trained solely on the original labeled examples with that of a classifier trained on both the original and synthetic examples generated and selected using our method. The performance evaluation was done using a separate held-out test set. 

The QWK score \cite{fleiss1973equivalence} measures the agreement between predicted and actual classifications, often used for ordinal or categorical data. It considers both actual and expected agreement, factors in the degree of disagreement using a weighted matrix, and handles unbalanced data well. QWK assigns greater importance to closer matches between categories, penalizing larger discrepancies more severely. The score ranges from -1 to 1: higher values imply better agreement, 0 suggests chance-level agreement and negative values indicate worse-than-chance agreement. QWK helps assess how well predictions match true values, considering both the quality of agreement and the potential for random agreement.

\section{Results}
\paragraph{GAN training}
The best FID scores obtained in a training run for each GAN model are summarized in Table \ref{GAN performance}. According to our experiments, transfer learning enables better FID scores. For this work, we employed class-specific GANs, which demonstrate better FID scores.

\begin{table}[ht]
\centering
\caption{Best FID scores from class-specific (with transfer learning) and class-conditional (without transfer learning) for each dataset}
\label{GAN performance}
\begin{tabular}{|lccccc|}
\hline
\multicolumn{6}{|c|}{\textbf{Best FID scores for conditional and specific GAN}} \\ \hline
\multicolumn{1}{|l|}{\textbf{}} & \multicolumn{1}{l|}{\textbf{Conditional}} & \multicolumn{4}{c|}{\textbf{Specific}} \\ \hline
\multicolumn{1}{|l|}{\textbf{Dataset}} & \multicolumn{1}{c|}{} & \multicolumn{1}{c|}{MES 0} & \multicolumn{1}{c|}{MES 1} & \multicolumn{1}{c|}{MES 2} & MES 3 \\ \hline
\multicolumn{1}{|l|}{\textbf{Subset 50}} & \multicolumn{1}{c|}{154.8} & \multicolumn{1}{c|}{129.7} & \multicolumn{1}{c|}{110.7} & \multicolumn{1}{c|}{100.6} & 115.5 \\ \hline
\end{tabular}
\end{table}

\paragraph {Synthetic endoscopy image generation} 
Synthetic image generation component of this work includes class-specific GAN training for the dataset. A different GAN model for each class were trained. GAN training for all models were concluded at the 5 million image mark, i.e., GAN discriminator model had seen 5 million synthetic images that were generated by the generator model. The Fréchet Inception Distance (FID) metrics for all GAN models were monitored and we chose the lowest (best) FID score model.
   			
\paragraph {Comparison of active learning methods} 

In the literature, there are two primary categories of methods: diversity-based and uncertainty-based approaches. In our experimental analysis, we delved into uncertainty-based methods such as Entropy and Margin, as well as a diversity-based method known as Coreset. Furthermore, we introduced the concept of `neighbour margin', which not only assesses the margin between the top two predictions, but also takes into account the ordinal proximity of their labels. We have demonstrated that our neighbour margin method outperforms the techniques discussed in the literature when applied to our dataset. It demonstrates superior performance compared to the gains achieved through the inclusion of up to twice the original image quantity per class.

\begin{figure}[ht]
    \centering
    \begin{tabular}{c c c c}
        \textbf{{MES 0}} & \textbf{{MES 1}} & \textbf{{MES 2}} & \textbf{{MES 3}}\\
        healthy & mild & moderate & severe\\
        \includegraphics[width=0.10\textwidth]{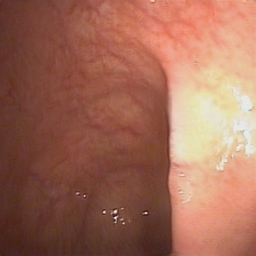} & 
        \includegraphics[width=0.10\textwidth]{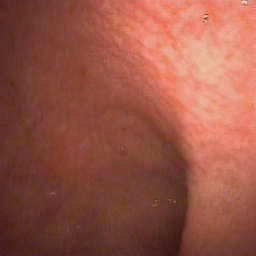} &
        \includegraphics[width=0.10\textwidth]{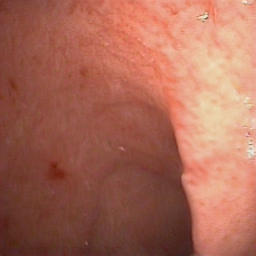} & 
        \includegraphics[width=0.10\textwidth]{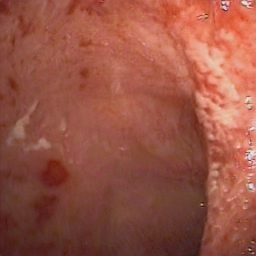}
        
    \end{tabular}
    \caption{Synthetic Endoscopy Images generated with StyleGAN2 with adaptive discriminator augmentation (ADA) method from the same seed.}
    \label{fig:smayo_images}
\end{figure}

The study employed a GAN to create images representing different levels of MES from 0 to 3. These images were generated using random noise input to the generator network, resulting in realistic colonoscopy visuals. The focus was solely on the generator network within the GAN architecture. The synthesized images, produced with the same random seed, are showcased in Figure \ref{fig:smayo_images}. The images clearly exhibit increasing inflammation as expected from increasing MES, effectively demonstrating the impact of UC on visual representations. This GAN-based approach effectively captures the visual progression of UC-induced inflammation according to MES levels.

\paragraph{MES Classification}

\begin{table}[ht]
\begin{center}
\caption{Performance results in percentage for varying dataset sizes}
\begin{tabular}{|l|c|c|c|c|c|}
\hline
Set & F1   & QWK  & Acc  & Precision & Recall  \\ \hline
50  & 54.9 & 68.1 & 59.1 & 53.9   & 57.6 \\ \hline
100 & 58.9 & 74.2 & 63.9 & 58.4   & 61.0 \\ \hline
150 & 60.6 & 76.0 & 66.0 & 60.6   & 63.4 \\ \hline
200 & 62.5 & 77.9 & 67.4 & 61.6   & 64.5 \\ \hline
250 & 63.2 & 78.3 & 67.9 & 61.8   & 65.7 \\ \hline
300 & 63.9 & 79.2 & 69.0 & 62.7   & 66.2 \\ \hline
350 & 64.0 & 79.2 & 69.8 & 62.4   & 66.7 \\ \hline
400 & 65.4 & 80.5 & 71.0 & 63.8   & 67.9 \\ \hline
450 & 65.6 & 80.1 & 70.3 & 64.1   & 68.0 \\ \hline
500 & 67.0 & 80.9 & 71.0 & 65.3   & 69.6 \\ \hline
\end{tabular}
\label{tab:saturation-real}
\end{center}
\end{table}

Performance of the MES classification was evaluated using various metrics including F1 score, QWK score, accuracy, precision, and recall. This evaluation was conducted on different subsets, each containing a specific number of images per class. As anticipated, the classification model's performance demonstrated improvement as the number of training images increased. However, this improvement reached a saturation point for the QWK score at around 300 images per class. Further additions of real images beyond this point did not yield further performance improvements as observed in Table \ref{tab:saturation-real}. 

\begin{table}[ht]
\begin{center}

\caption{MES classification performance in percentage of the mean performance metrics for each active learning method}
\begin{tabular}{|l|c|c|c|c|c|}
\hline
\multicolumn{1}{|c|}{}                                             & F1               & QWK              & Acc              & Precision        & Recall           \\ \hline
Baseline                                                           & 54.9          & 68.1          & 59.1          & 53.9          & 57.6          \\ \hline
Random                                                             & 59.8          & 73.5          & 64.3          & 58.9          & 61.9         \\ \hline
Entropy                                                            & 59.8          & 74.1          & 65.0          & 58.7          & 62.0          \\ \hline
Margin                                                             & 60.2          & 74.2          & 65.3          & 59.1          & 62.4          \\ \hline
Coreset                                                            & 60.4          & 74.2          & 64.8          & \textbf{59.6} & 62.1          \\ \hline
\begin{tabular}[c]{@{}l@{}}Neighbour \\ Margin (ours)\end{tabular} & \textbf{60.5} & \textbf{74.5} & \textbf{65.4} & 59.3          & \textbf{62.6} \\ \hline
\end{tabular}
\label{tab:Summary Performance Table}
\end{center}
\end{table}

The summary of the MES classification performance using various active learning methods is presented in Table \ref{tab:Summary Performance Table}. The baseline indicates the performance using 50 images per class, employing a similar approach in \cite{gorkem_p099}. 

In response to sample similarity arising from GAN training with limited data, Coreset algorithm was employed to eliminate similar samples and establish a diverse subset. Subsequently, diverse samples were systematically selected from this subset using different strategies including random, margin, entropy, coreset, and neighbour margin techniques. These approaches yield progressively larger subsets, each capturing varied data representations. Ultimately, active learning methods enhance the utility of generated samples from a GAN model trained with limited data by curating subsets with improved variety.

As opposed to the extensive optimization needed for generative data augmentation, we have introduced active learning methods that choose and utilize synthetic images to enhance data augmentation. Our results illustrate the superiority of nearly all active learning methods over the random selection approach. Notably, our proposed method, neighbour margin, emerges as the most effective among all the tested methods.

Classification performance of the model increased steadily with the increasing number of real images used in training. As shown in Table. \ref{tab:saturation-real}, the mean QWK score performance of the classification network was 68.1\% with 50 real images per class, which increased up to 80.9\% with 500 real images per class. The improvement of QWK classification score saturated around 200 per class subset with addition of extra real images improved the QWK score only less than 1 percentage point.

The saturation experiment was also valuable to determine the baseline classification performance for all subsets. The classification performance of a deep learning model trained only with real images was later used as a baseline. The later experiments aimed to increase the performance of the classification model with the addition of synthetic images.

\section{Discussions}

Finding the best truncation value for image synthesis and the best number of synthetic images is an exhaustive process and it is not feasible for practical applications. We have done an extensive grid search for various truncation values $T=\{0.5,0.6,0.7,0.8,0.9,1.0,1.2,1.5,2.0\}$ and number of synthetic images from $1\times$ to $10\times$ the original real images per class. And, we have found that different dataset sizes require different number of synthetic images generated with different truncation values. This conflicting and non-converging result have led us to believe an adaptive way of actively selecting influential images could be beneficial for generative data augmentation.

Our results showed that using synthetic examples in conjunction with active learning led to better classification performance compared to using only the original labeled examples. Specifically, the classifier trained on synthetic examples had higher QWK, accuracy, precision, and recall than the classifier trained on only the original labeled examples. These results demonstrate the effectiveness of our proposed method for UC classification using deep learning models. 

In safety-critical domains, generating accurate annotations necessitates a unanimous consensus from a committee of expert annotators. This stringent requirement poses challenges to annotating extra data and expanding the dataset. Furthermore, our findings show that the application of generative data augmentation in conjunction with active learning methodologies yields superior results in comparison to incorporating additional real samples. As a results, augmenting real images with synthetic data emerges as a feasible and practical alternative.

\section{Conclusions}

Severity assessment of UC involves an unanimous decision from a committee of physicians known as Mayo Endoscopic Scoring (MES). This process can be automated or aided by employing deep learning methods. However, deep learning methods require data to work effectively. Due to the cost of acquisition of data, especially in medical domain, it is imperative to work with limited data. Data augmentation methods are proven to be an effective and robust way of improving deep learning model performance with limited data. Generative data augmentation is a novel way of improving model performance with synthetic data. Since the generative data augmentation methods trained with limited data tend to generate similar examples, the dataset needs to be further reduced by diversity algorithm to overcome this problem.

We have shown that deep learning based classification performance can be improved with the generative data augmentation employing synthetic data. Extensive hyperparameter tuning and empirical experimentation is a resource-heavy and exhaustive process. Therefore, we generated a large synthetic data pool and sampled diverse images using coreset algorithm. Active learning is a method to find a good combination of samples from a synthetic dataset. Therefore, we applied active learning algorithms on the sampled diverse subset to further improve MES classification performance. In our work active learning methods of entropy, coreset and margin from the literature and our proposed method of neighbour margin achieved better results than random sample selection. Furthermore, the proposed neighbour margin method achieved the best results among the other alternatives.

We aimed to select the uncertain samples at the neighbour class boundaries, considering the severity of the disease is ordinal. Due to this ordinal structure, we anticipate a higher likelihood of estimates being confused between adjacent classes. The confusion of distant classes instead of neighboring classes actually increases the probability of these synthetic samples being outliers. For this reason, the neighbour margin method promotes selecting synthetic samples at class boundaries by considering the margin values between neighboring classes.

One important outcome of this work is the demonstration of addition of synthetic data, with active learning strategies, can achieve better results than addition of real data over the baseline. This is especially valuable in the medical domain as the cost of new real data can be a limiting factor whereas deep learning methods, such as approaches we have employed in this work, can achieve improved performance with larger datasets. Our results showed that even with limited data, we can achieve improvements over the baseline with generative data augmentation enhanced with active learning method. Finally, our proposed method, neighbour margin, achieves the best results among other active learning methods.

\section*{Acknowledgement}

 The numerical calculations reported in this paper were fully performed at TUBITAK ULAKBIM, High Performance and Grid Computing Center (TRUBA)

\bibliographystyle{plain}
\bibliography{bibfile}


\end{document}